\documentclass[wsdraft]{ws-procs961x669}  
 \begin{document}
\title{On large mass $\gamma -\gamma$ and $\gamma ~-$ meson photoproduction }
\author{L. Szymanowski$^*$}
\address{NCBJ, 02-093 Warsaw, Poland\\
$^*$E-mail: Lech.Szymanowski@ncbj.gov.pl }
\author{B. Pire}
\address{CPHT, CNRS, \'Ecole polytechnique, I.P. Paris, 91128-Palaiseau, France}
\author{S. Wallon}
\address{LPT, CNRS, Univ. Paris-Sud, Universit\'e Paris-Saclay, 91405, Orsay, France {\em \&} \\
Sorbonne Universit\'e, Facult\'e de Physique, 4 place Jussieu, 75252 Paris Cedex 05, France}
\begin{abstract}
Enlarging the set of hard exclusive reactions to be studied in the framework of QCD collinear
factorization opens new possibilities to access generalized parton distributions (GPDs). We
studied the photoproduction of a large invariant mass photon-photon or photon-meson pair in
the generalized Bjorken regime which may be accessible both at JLab and  at the EIC.
\end{abstract}
\bodymatter
\section{Introduction}\label{aba:sec1}
Deeply virtual Compton scattering (DVCS) has proven to be a promising tool to study the three
dimensional arrangement of quarks and gluons in the nucleon \cite{Kumericki:2016ehc}. The
crossed reaction, the photoproduction of a timelike highly virtual photon which materializes
in a large invariant mass lepton pair (dubbed TCS for timelike Compton scattering) is under
study at JLab. Its amplitude shares many features with the DVCS amplitude
\cite{Berger:2001xd} but with significant and interesting differences \cite{Pire:2011st,
Muller:2012yq, Moutarde:2013qs} due to the analytic behavior in the large scale $Q^2$
measuring the virtuality of the incoming ($q^2 =-Q^2$) or outgoing  ($q^2 =+Q^2$) photon. In
order to enlarge the set of experimental data allowing the extraction of GPDs, we studied the
generalization of TCS to the case of the photoproduction of large invariant mass photon-
photon and photon-meson pairs. Although factorization of GPDs from a perturbatively calculable coefficient function has not yet been proven for these processes, they are a natural 
extension of the current picture in the framework of collinear QCD factorization.
\section{$\gamma N \to \gamma \gamma N'$}
\begin{figure}
\includegraphics[width=1.7in]{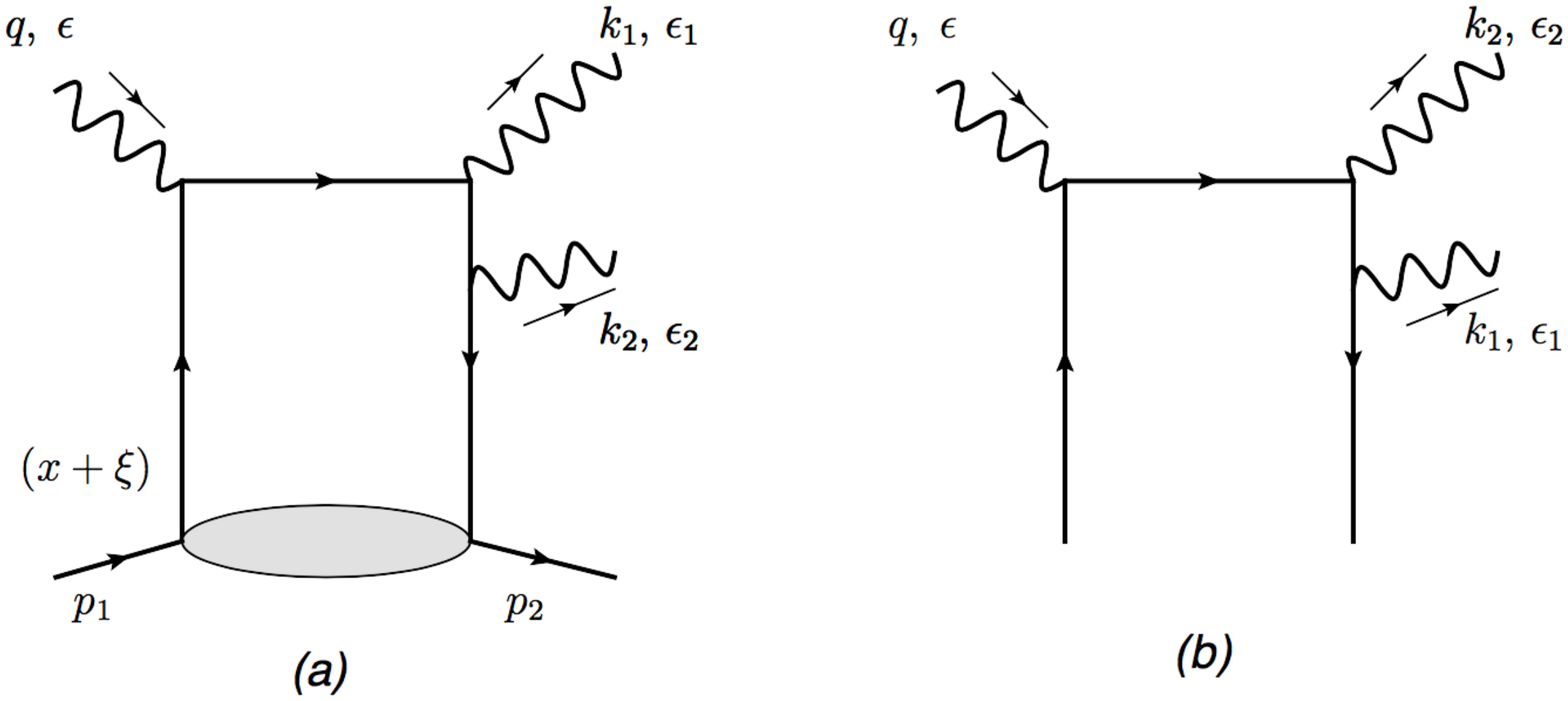}\includegraphics[width=1.7in]{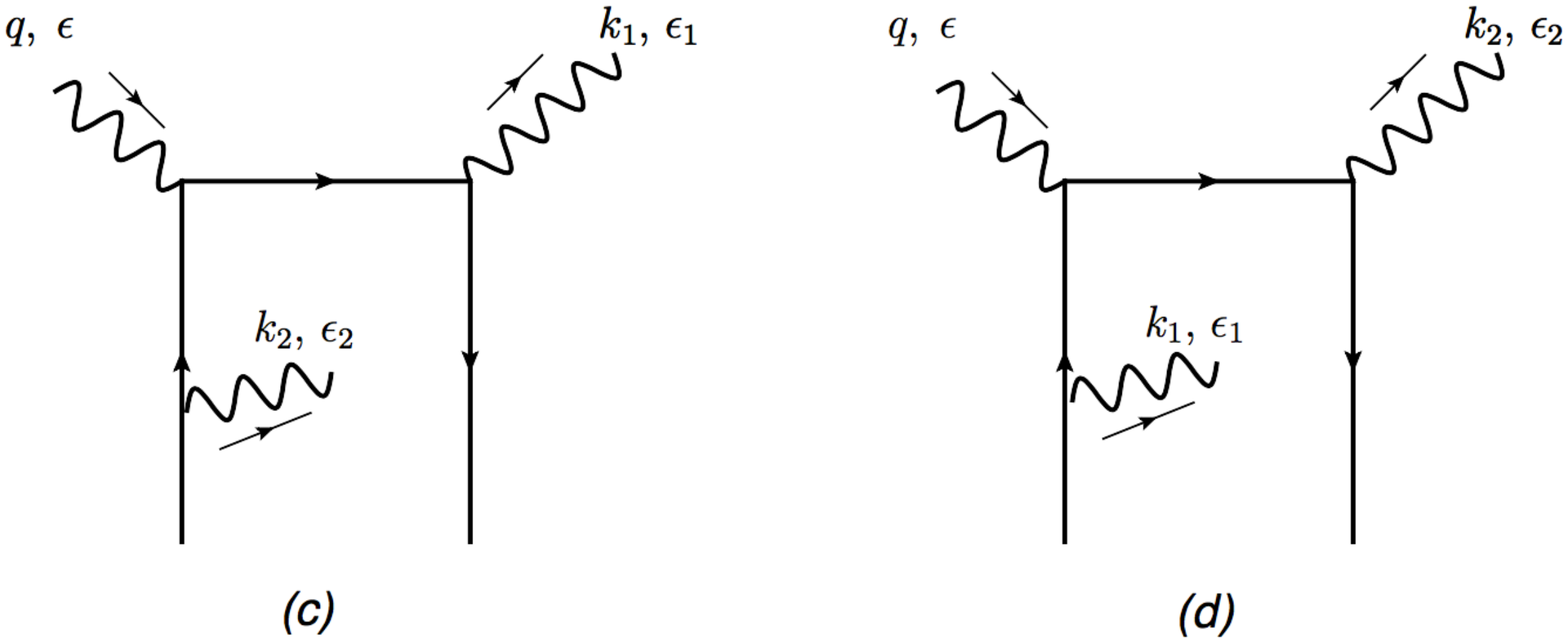}\includegraphics[width=1.7in]{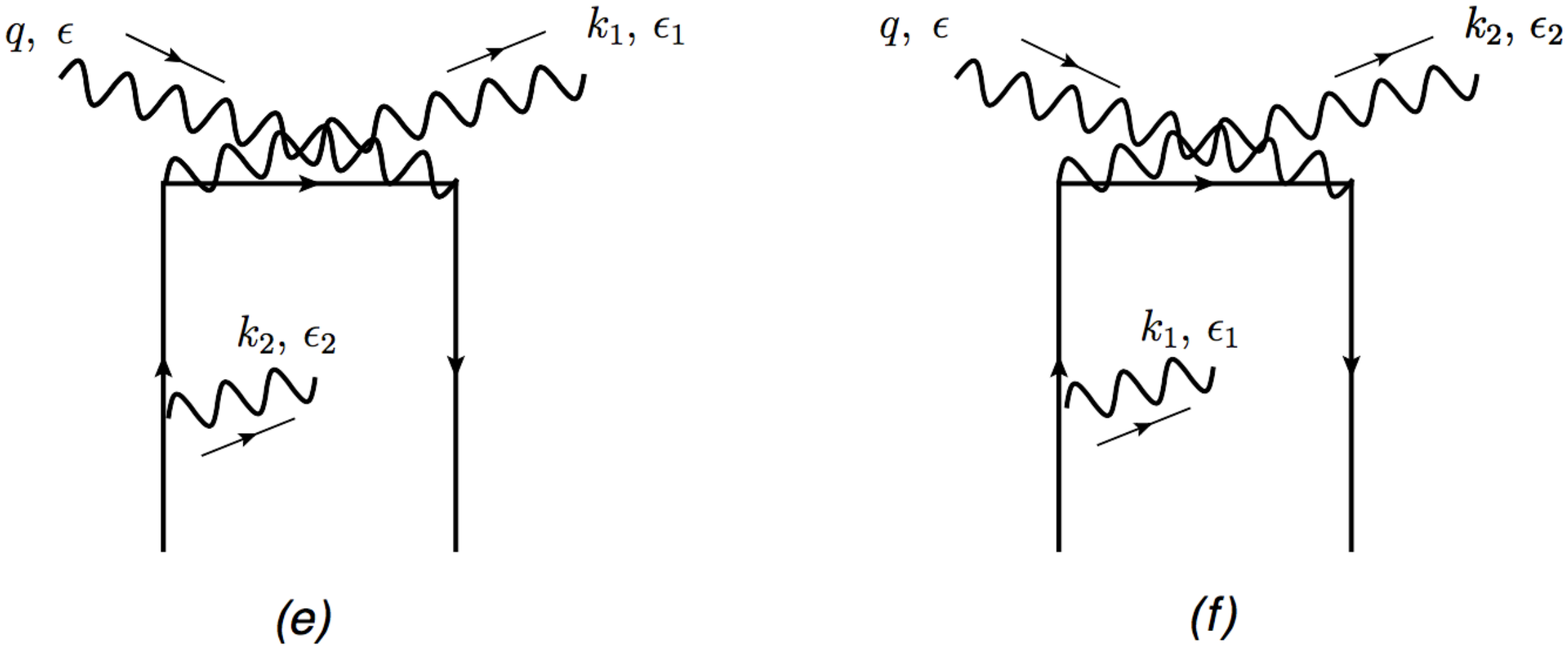}
\caption{Diagrams contributing to the coefficient function for $\gamma\gamma$ production at
the Born level.}
\label{aba:fig1}
\end{figure}
\begin{figure}
\center
\includegraphics[width=2.5in]{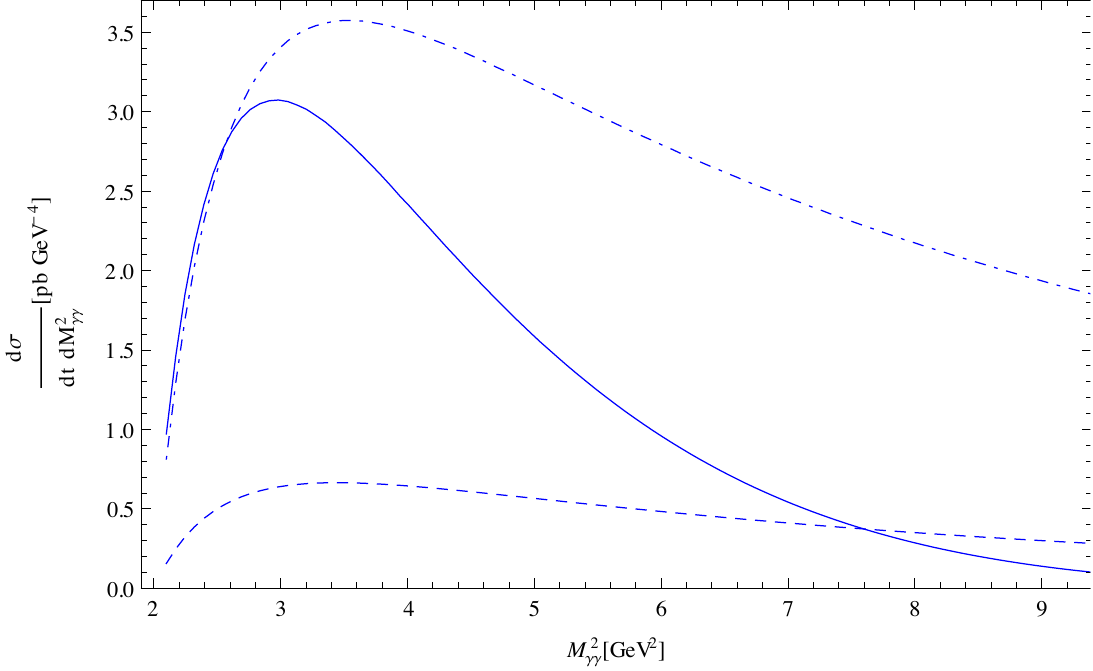}
\caption{$M_{\gamma\gamma}^2$ dependence of the unpolarized differential cross-section for the photoproduction of a diphoton on a proton (left panel) or neutron (right panel) at $t=t_{min}$ and $s_{\gamma N}= 20$ (resp. $100, 10^6$) GeV$^2$ (full, resp. dashed, dash-dotted multiplied by $10^5$).}
\label{gaga}
\end{figure}

The photoproduction of a photon pair \cite{Pedrak:2017cpp}
 shares with DVCS and TCS  the nice feature to be a purely electromagnetic amplitude at
the Born level. Charge parity however selects a complementary set of GPDs, namely the charge
parity - odd GPDs related to the valence part of quark PDFs, with no contribution from the
gluon GPDs. The analytic form of the Born amplitude calculated from the graphs shown on Fig.~\ref{aba:fig1} is very peculiar since the coefficient
function turns out to be proportional to $\delta(x\pm \xi)$ leading through the usual
momentum fraction integration to a scattering  amplitude  proportional to the GPDs taken at
the border values $x=\pm\xi$. For illustration, Fig.~\ref{gaga} displays the  
diphoton invariant squared mass dependence of the unpolarized differential cross section on a proton at $t = t_{min}$ and $s_{\gamma N}= 20$ (resp. $100, 10^6$) GeV$^2$ (full, resp. dashed, dash-dotted multiplied by $10^5$). 

\section{$\gamma N \to \gamma \rho N'$ : the quest for transversity GPDs}
The photoproduction of a $\gamma \rho$ pair \cite{ Boussarie:2016qop}
has the rare feature of being sensitive to chiral-odd transversity quark GPDs at the leading
twist level, because of the leading twist chiral-odd distribution amplitude of the
transversely polarized vector meson. Indeed, except for higher twist amplitudes which suffer
from end-point divergences and heavy meson neutrino production amplitudes
\cite{Pire:2015iza,Pire:2017lfj}  which may be difficult to measure, one needs exclusive
processes with more particules in the final state to access transversity GPDs
\cite{Ivanov:2002jj, Enberg:2006he, Beiyad:2010cxa, Cosyn:2019eeg}.

\begin{figure}
\includegraphics[width=4.2in]{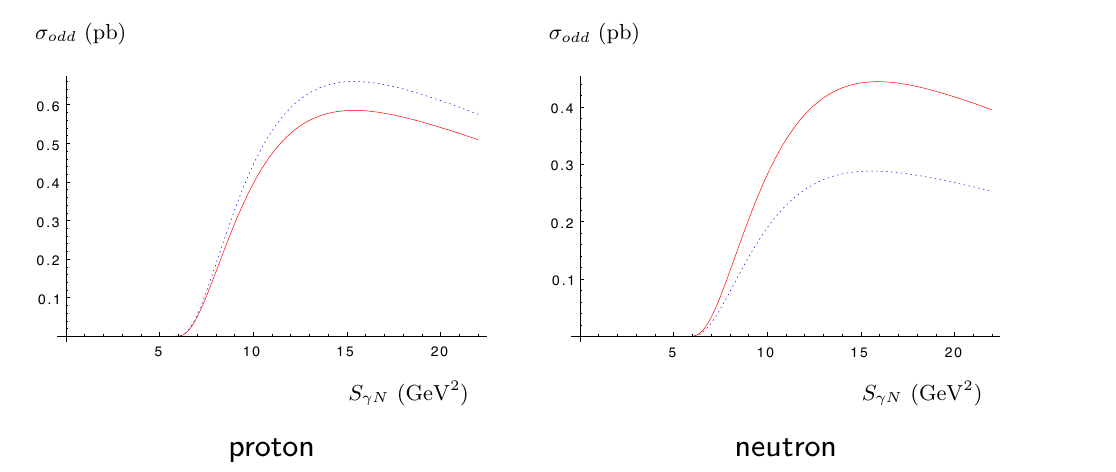} 
\caption{Energy dependence of the integrated cross section for a photon and a   transversely polarized $\rho$ meson production, on a proton (left panel) or neutron (right panel) target. The $\gamma \rho$ pair is required to have an invariant  mass squared larger than $2$ GeV$^2$.
The solid red and dashed blue curves correspond to different parametrization of the transversity GPDs.}
\label{garho}
\end{figure}

We show on Fig.~\ref{garho} the cross section for the production of a transversely polarized $\rho$ in conjunction with a photon, on a proton or a neutron target. The curves show the  sensitivity to the transversity GPD parametrization. Cross sections are sufficiently high for the process to be measurable at JLab \cite{Boussarie:2016qop}.

\section{$\gamma N \to \gamma \pi N'$}

\begin{figure}[h]
\center \includegraphics[width=4in]{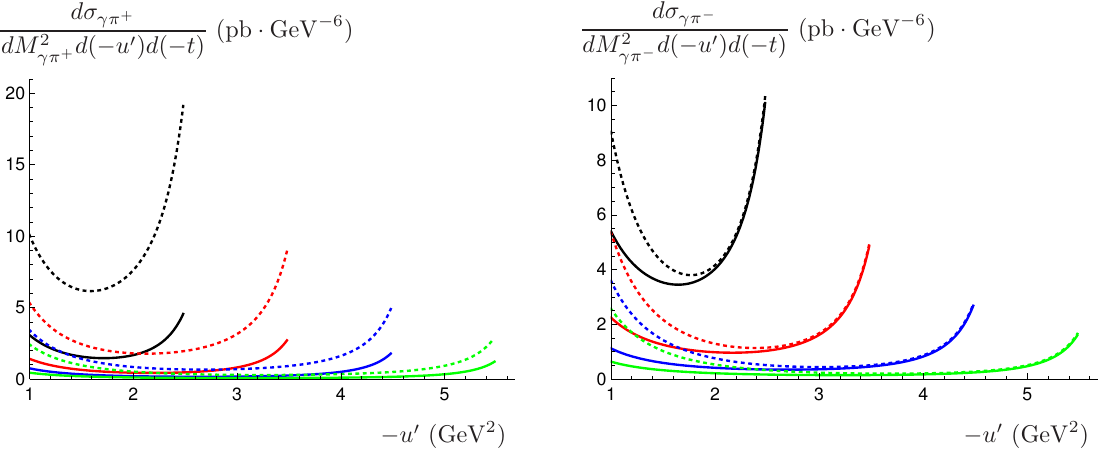}
\caption{ Left panel : the differential cross section for $\gamma \pi^+$  production on a proton target  at $s_{\gamma N} = 20$ GeV$^2$, $t=t_{min}$, and $M^2_{\gamma \pi}= 3$ (resp.$4,5,6$) GeV$^2$ for the black (resp. red, blue, green) curves. The solid and dashed  curves correspond to two different parametrization of the axial GPDs. Right panel : the same curves  for $\gamma \pi^-$  production on a neutron target.
}
\label{gapi}
\end{figure}

Since deep electroproduction of a $\pi$ meson has been shown to resist at moderate $Q^2$ to
leading twist dominance in the factorization framework, it has been tempting to  put the
blame on the peculiar chiral behavior of the higher twist (chiral-odd) pion DA as compared
with the leading twist (chiral even) pion DA. However, the dominance of higher twist contributions may not be a common feature of all exclusive amplitudes involving the pion DA. To check this idea, we propose \cite{Duplancic:2018bum}
 to study  the related process  $\gamma N \to \gamma \pi N'$
where the same pion DAs appear. It turns out that the axial nature of the pion leading twist DA infers a high sensitivity of the amplitudes to the axial GPDs $\tilde H(x,\xi,t)$ as shown on Fig.~\ref{gapi} where the cross sections for the reaction $\gamma p \to \gamma \pi^+ n$ and $\gamma n \to \gamma \pi ^-p$ are displayed for two different sets of axial GPDs. The rates are of the same order as for the $\gamma \rho$ case and we thus expect these reactions to be measurable at JLab.

\section{Conclusions}
The processes discussed here, because of the absence of gluon and sea quark contributions are not enhanced at high photon energy (or small skewness $\xi$) and they are thus more accessible at JLab than at EIC. However, since a high energy electron beam is also an intense source of medium energy quasi real photons ($q^2\approx 0$), with fractions of energy $y = \frac{q.p}{k.p}=0(10^{-3})$, ($k$ and $p$ being the initial electron  and nucleon momenta), one may expect the $\gamma \gamma$ and $\gamma \rho_L$ channels to be accessible at moderate values of $s_{\gamma N}$. Prospects at higher values of $s_{\gamma N}$ (and smaller values of the skewness $\xi$) are brighter for the $\gamma \pi^0$ channel which  benefits from the contributions of small $\xi$ sea-quark and gluon GPDs. 
\section*{Acknowledgments}
We acknowledge the collaboration of R. Boussarie, G.~Duplancic, K.~Passek-Kumericki, A. Pedrak and J. Wagner for the works reported here. L. S. is supported by the grant 2017/26/M/ST2/01074 of the National Science Center in Poland. He thanks the French LABEX P2IO and the French GDR QCD for support. 

\end{document}